\documentstyle[12pt,epsfig]{article}
\def    \be             {\begin{equation}}
\def    \ee             {\end{equation}} 
\def    \ba             {\begin{eqnarray}}
\def    \ea             {\end{eqnarray}} 

\def    \=              {\;=\;} 
\def    \frac           #1#2{{#1 \over #2}}

\def    \ie             {{\em i.e.\/} }

\def \jpsi {\mbox{$\Psi$}}

\def \as   {\mbox{$\alpha_s$}}

\def \asopi {\frac{\alpha_s}{\pi}}
\def \oacube {\mbox{${\cal O}(\alpha_s^3)$}}

\def \pt   {\mbox{$p_t$}}                
                
\newcommand\sss{\scriptscriptstyle}
\newcommand\mur{\mbox{$\mu_{\sss R}$}}  
\newcommand\muf{\mbox{$\mu_{\sss F}$}}  
\def \to   {\mbox{$\rightarrow$}}

\newcommand{\ccaption}[2]{
  \begin{center}
    \parbox{0.85\textwidth}{
      \caption[#1]{\small\it {#2}}}
  \end{center}    }
 
\def\qed{\hbox{${\vcenter{\vbox{                        %HOLLOW SQUARE
   \hrule height 0.4pt\hbox{\vrule width 0.4pt height 6pt
   \kern5pt\vrule width 0.4pt}\hrule height 0.4pt}}}$}}

\begin{document}

%\copyrightheading{}                     %{Vol. 0, No. 0 (1993) 000--000}
 
\begin{titlepage}
\nopagebreak
{\flushright{
        \begin{minipage}{4cm}
        CERN-TH/96-293\\
        hep-ph/9610364\\
        \end{minipage}        }

}
\vfill                        
\begin{center}
{\LARGE { \bf \sc NLO Quarkonium Production\\[0.3cm]in Hadronic Collisions}}
\footnote{To appear in the                          
Proceedings of the Quarkonium Physics Workshop, University of Illinois,
Chicago, June 13--15, 1996.}
\vfill                                
\vskip .5cm                               
{\bf Michelangelo L. MANGANO
\footnote{Presenting author. On leave of absence from INFN, Pisa, Italy.
E-mail: {\tt mlm@vxcern.cern.ch}} and Andrea PETRELLI
\footnote{Permanent address: Dipartimento di Fisica dell'Universit\'a and
INFN, Pisa, Italy.
E-mail: {\tt petrelli@mail.cern.ch}}
}            
                   
\vskip .3cm                                      
{CERN, Theoretical Physics Division,\\
1211 Geneva 23, Switzerland}\\

\end{center}                                                      
\nopagebreak
\vfill                                
%\vskip 3cm                                               
\begin{abstract}
We present some preliminary results on the next-to-leading order 
calculation in QCD of quarkonium production cross
sections in hadronic collisions. We will show that the NLO total cross sections
for $P$-wave states produced at high energy are not reliable, due to the
appearance of  very large and negative contributions. We also discuss some
issues related to the structure of final states in colour-octet production and
to high-\pt\ fragmentation.
\end{abstract}                                                               
\vskip 1cm
CERN-TH/96-293 \hfill \\
October 1996 \hfill
\vfill       
\end{titlepage}
       
\section{Introduction}          %) A SECTION HEADING
The production of quarkonium states in hadronic collisions has recently
attracted a lot of interest in the theoretical community, as the contributions
to this Workshop confirm. Most
of the studies done so far have concentrated on key issues such as whether the
colour-octet mechanism~\cite{Braaten96} 
can indeed explain both Tevatron and fixed-target data, 
and on trying to identify the most direct and distinct signatures of it.
Applications to production of quarkonium at LEP and HERA have also been
considered. In general, I believe it is 
fair to say that the field is still in its infancy. The theoretical predictions
require the introduction of several new nonperturbative parameters~\cite{bbl}
to describe
the values of colour-singlet and colour-octet
operators on different states, and the available data can only in part fix
these values. 
In absence of exactly known relations between
the values of the different nonperturbative matrix elements, connecting for
example their values on 1S states to their values on 2S or P states, it is
difficult to reduce the number of independent parameters. This  increases the
number of separate measurements needed to fix them, and reduces the set of
data available to test the predictions of the theory. 
While the use of nonperturbative matrix elements extracted from
the Tevatron data~\cite{Braaten95} 
provides an acceptable description of the fixed-target
data~\cite{Beneke96},
the naive use of the Tevatron fits at HERA significantly overestimates the
yield of observed \jpsi\ in the region $z\to 1$~\cite{Cacciari96}.
Whether this is a real problem of the current theory,             
or whether it is just a result of our incomplete understanding of it, is a
question that still waits an answer. 

With so many very basic open questions, I decided nevertheless to concentrate 
in this presentation on the                                                   
subject of higher-order QCD corrections. Given the size of the current 
theoretical uncertainties, I consider this subject perhaps academic. 
For example~\cite{Rothstein96}, although there are arguments that the right
mass parameter to be used in the perturbative evaluation of the  short-distance
matrix elements and in the phase space constraints 
should be $2m_Q$ (\ie twice the heavy-quark mass), it is not clear that this
prescription properly reflects the correct nonperturbative result, where one
expects that, at least  for the determination of the phase space boundary,
the quarkonium mass should be used. The                                   
two choices lead to results which differ by large factors. Undertaking the
task of calculating next-to-leading order (NLO) corrections to the LO results
seems therefore a bit premature, and the hope that the inclusion of NLO effects
could help making the predictions more accurate is in my view, today, not
supported by solid evidence. The reason why I am interested in higher-order
corrections is that hopefully they will help learning more about the structure
of perturbation theory for quarkonium production in both the colour-singlet and
colour-octet models. It is a well known fact that NLO corrections to charmonium
decay widths are very large. It is important to remark that their size is not
just a consequence of the large value of \as\ evaluated at the charm mass
scale: it is mostly the result of very large coefficients which multiply
\as\ in the radiative corrections.
It is interesting to see what happens of these large coefficients when we study
charmonium production at NLO. Should the size of NLO corrections be very large,
the whole exercise of extracting the value of nonperturbative parameters from a
comparison of LO matrix elements with data, and the use of these parameters to
perform predictions for different observables, would clearly have less chances
of producing reasonable results. This is because NLO corrections to different
observables might be in principle very different in size.                 
I would not be surprised if this were part of the solution to the puzzle
uncovered by the application of the colour-octet model to the HERA data.
                                                                        
In the present talk I will first of all 
briefly illustrate the technique we used to evaluate the NLO cross sections.
This technique makes use of dimensional regularization, but uses universal
properties of soft and collinear singularities to avoid the need of calculating
the real-emission cross sections in $D$ dimensions. This technique was first
introduced for hadronic processes in ref.~\cite{Mele91}, and in the specific
context of heavy quark production in ref.~\cite{Mangano92}.
The full details of the NLO quarkonium calculation, including explicit results
for colour-singlet $^1S_0^{[1]}$ and $^3P_{0,2}^{[1]}$  states and for several
colour-octet states ($^1S_0^{[8]}$, $^3S_1^{[8]}$, $^3P_{0,2}^{[8]}$),
will be documented in a forthcoming publication~\cite{Inprogress}.    
The NLO cross sections for the $^1S_0^{[1]}$ state have already appeared in the
literature, in the papers of K\"uhn and Mirkes~\cite{Kuehn93}
and Schuler~\cite{Schuler94}. Our results and theirs for this channel
are in full agreement.
                                                                           
Next I will show some numerical results. For simplicity, I will just confine
myself to the case of $^3P_{0,2}^{[1]}$ states, studied 
at fixed-target and at collider
energies. While the results at fixed target are extremely encouraging,
displaying relatively small $K$-factors and a significant reduction of the 
scale dependence, the results at collider energies are very worrisome. In
short, the radiative corrections turn out to be extremely large and negative,
so as to leave us with negative total cross sections. The origin of this
problem will be discussed in some detail.

To conclude, I will make a few remarks on the issue of understanding the full
structure of the final states produced in conjunction with quarkonium, with
particular reference to the production via fragmentation of a gluon jet.

The style of this written contribution is very informal, trying to convene to
the reader who did not attend the wonderful spirit of this
Workshop.  I wish to thank the organizers, the session chairpersons and, most
of all, the participants, for the great atmosphere in which the Workshop took
place. I look forward to more opportunities like this to openly discuss future
progress in the field!

\section{The Structure of Quarkonium Total Cross Sections}
We are interested here in the process $h_1 \; h_2 \to O + X$,
where $h_{1,2}$ are arbitrary hadrons and $O$ is a quarkonium state. We will
limit ourselves to states for which the lowest-order process $gg\to O$ is
non-zero. In this way the \oacube\ contributions represent genuine NLO effects.
These involve the evaluation of the
virtual corrections to the $ggO$ vertex, in addition to the real emission
processes $gg\to Og$, $qg\to Oq$ and $q\bar q \to gO$.
We will consider these processes separately. 

The contribution of the $gg$ process to the total cross section gives rise to
collinear and soft singularities. These can be regulated by working in
dimensional regularization, giving rise to poles in $1/\epsilon$: double
poles for the 
leading soft singularities, and single poles for the sub-leading soft and for
the collinear singularities. The soft singularities are absorbed by similar
singularities present in the virtual corrections to the $ggO$ vertex, leaving
finite terms which contribute to the processes with Born-like kinematics $gg\to
O$. The collinear singularities are absorbed into a redefinition of the 
initial-state parton densities, according to the standard procedure of          
factorization of the initial-state mass singularities. The residual finite
contributions correspond to purely inelastic processes $gg\to Og$, regulated at
the boundary of phase space (namely in the soft and collinear region) by an
appropriate ``+'' prescription.                                   
Given the simplicity of the
kinematics of the LO process, and given the universal character of soft and
collinear emission, it can be shown~\cite{Inprogress} that 
the structure of the NLO partonic cross section is given in
general\footnote{Although I will only present here results for P-waves, the
structure of the cross sections is exactly the same for the production of other
states, regardless of the $^SL_J$ quantum numbers.} by the following
expression:                                                
\ba        
	\sigma_{NLO}^{J,(gg)}(x) &-& \sigma_{Born}^{J,(gg)} \delta(1-x) \;=\;
       \asopi \sigma_{Born}^{J,(gg)}\; x \nonumber \\
   &\times  &\left\{  \delta(1-x) \left[ A_J - C_A\frac{\pi^2}{3} \right] 
  \;+\; F_J(x) \;-\; P_{gg}(x) \log{x}  \right.  \nonumber \\
  &+& \left. 4C_A \left[\left(\frac{1}{x}+x(1-x)-2\right)\log{(1-x)} \;+\; 
   \left[\frac{\log(1-x)}{1-x}\right]_+ \right] \right\}  \; ,
\ea                                                          
where $J=0,2$ is the total angular momentum of the $P$-wave state considered, 
$x=m^2/\hat s$, $m$ is the quarkonium mass, $C_A=N_c=3$ is the number of
colours, and $P_{gg}$ is the                           
Altarelli-Parisi splitting kernel. For simplicity I have put the factorization
(\muf) and renormalization scales (\mur) equal to $m$. This sets to zero some
universal terms proportional to $\log(\mur/m)$ or $\log(\muf/m)$. 
The factorization of collinear singularities was performed in the 
$\overline{MS}$ scheme. All of the dependence on the quarkonium state is
included in the numbers $A_J$ and in the functions $F_J(x)$. $A_J$ is related to
the finite part of the virtual corrections, defined by the equation:
\be                                                                
   	\sigma_{virt}^{J,(gg)} (x) \;=\;
     \asopi \sigma^{J,(gg)}_{Born,D} 
     \left( \frac{4\pi\mu^2}{\hat s}\right)^\epsilon \, \Gamma(1+\epsilon) \,
     \delta(1-x) \times                                                        
 \left( A_J + \frac{C_2}{\epsilon^2} + \frac{C_1}{\epsilon}\right) \; ,
\ee                                                                    
where $\sigma^{J,(gg)}_{Born,D}$ is the Born cross section in $D$ dimensions and
$C_1$ and $C_2$ are numerical coefficients independent of $\epsilon$. From an
explicit calculation, we get:
\ba \label{eq:B0}
  A_0 &=& C_F\left( -\frac{7}{3}+\frac{\pi^2}{4}\right) 
  + C_A \left( \frac{1}{3} + \frac{5}{12}\pi^2 \right) \\
    \label{eq:B2}
  A_2 &=& -4C_F
  + C_A \left( \frac{1}{3} + \frac{\pi^2}{6} +\frac{5}{3}\log 2 \right) \; .
\ea
The function $F_J(x)$ is given instead by the following relation:
\be  \label{eq:Fgg}
\asopi \sigma_{Born}^{J,(gg)} F_J^{(gg)} (x) \; = \;
\frac{1}{32\pi \hat{s}} \left( \frac{1}{1-x} \right)_+ 
 \; \int_{-1}^{1} dy \; \left(\frac{1}{1-y} \right)_+ 
  \overline{M^{J}}_{(4)}(x,y) \; ,
\ee                                                                            
where $y=\cos\theta$ is the cosine of the scattering angle in the hard process
CoM frame, and 
$\overline{M^J}_{(4)}(x,y)$ is related to the four-dimensional matrix
element squared for the $gg\to Og$ process by the following relation:
\be                                                                 
\overline{M^J}_{(4)}(x,y) \; = \; (1-x)^2 \, (1-y^2) M^J_{(4)}(x,y) \;.
\ee                                                                 
The explicit expressions will be reported in~\cite{Inprogress}.

The contribution of the $qg$ process to the total cross section only appears at
\oacube. It gives rise to 
singularities due to the collinear emission of the gluon entering
the hard scattering form the initial-state quark.
As before, these singularities are absorbed into a redefinition of the
parton densities, leaving a residual finite                     
contribution corresponding to the purely inelastic process $qg\to Og$.
Once more the universal character of 
collinear emission can be used to reduce the final result to a simple 
expression, given by:                                       
\be                            
	\sigma_{NLO}^{J,(qg)}(x) \;=\;
       \asopi \sigma_{Born}^{J,(gg)} \; x \times  
   \left[ F_J^{(qg)}(x) +P_{gq}(x)\log(1-x) +\frac{xC_F}{2}
    \right] \;.
\ee                                                    
All of the dependence on the quarkonium state in included in the function
$F_J^{(qg)}(x)$, which is defined by the analogous of eq.~\ref{eq:Fgg} and
will be given explicitly for the various states in ref.~\cite{Inprogress}.

The study of the $qq$ channel is interesting from the theoretical
point of view, since this channel
exhibits a singularity related to the quarkonium binding energy. This is the
analogous of the singularities found in the case of P-wave decays to 
$q\bar q g$. In dimensional regularization, we obtain:
\be                            
	\sigma_{NLO}^{J,(qq)}(x) \;=\;
       (2J+1) \left( -\frac{1}{\epsilon}\right) B \delta(1-x)
   + F_{J}^{(qq)}(x) \; ,
\ee                                                    
where $B$ is a constant factor, independent of $J$. The $1/\epsilon$ pole can
be removed by a renormalization of the $^3S_1^{[8]} \to ^3P_J^{[1]}$ transition
matrix element, and including the $q\bar q \to ^3S_1^{[8]} \to \;^3P_J^{[1]} X$
contribution to the cross section. 
Any reasonable value of the renormalized coupling will 
however make the numerical impact of the $qq$ channel totally negligible in
all experimental configurations of interest, except for
$\pi N\to \chi_b$ production at very low energy. We base this claim on a study
of the $qq$ production mechanism done using an IR cutoff in four dimensions.
In this scheme, the divergence is proportional to the logarithm of the binding
energy. Even with values as low as few MeV the $qq$ channel contribution
to $^3P_J$ production is overwhelmed by the $gg$ and $qg$ channels.
                                     
\section{Numerical Results}
\begin{figure}                                               
\centerline{\epsfig{figure=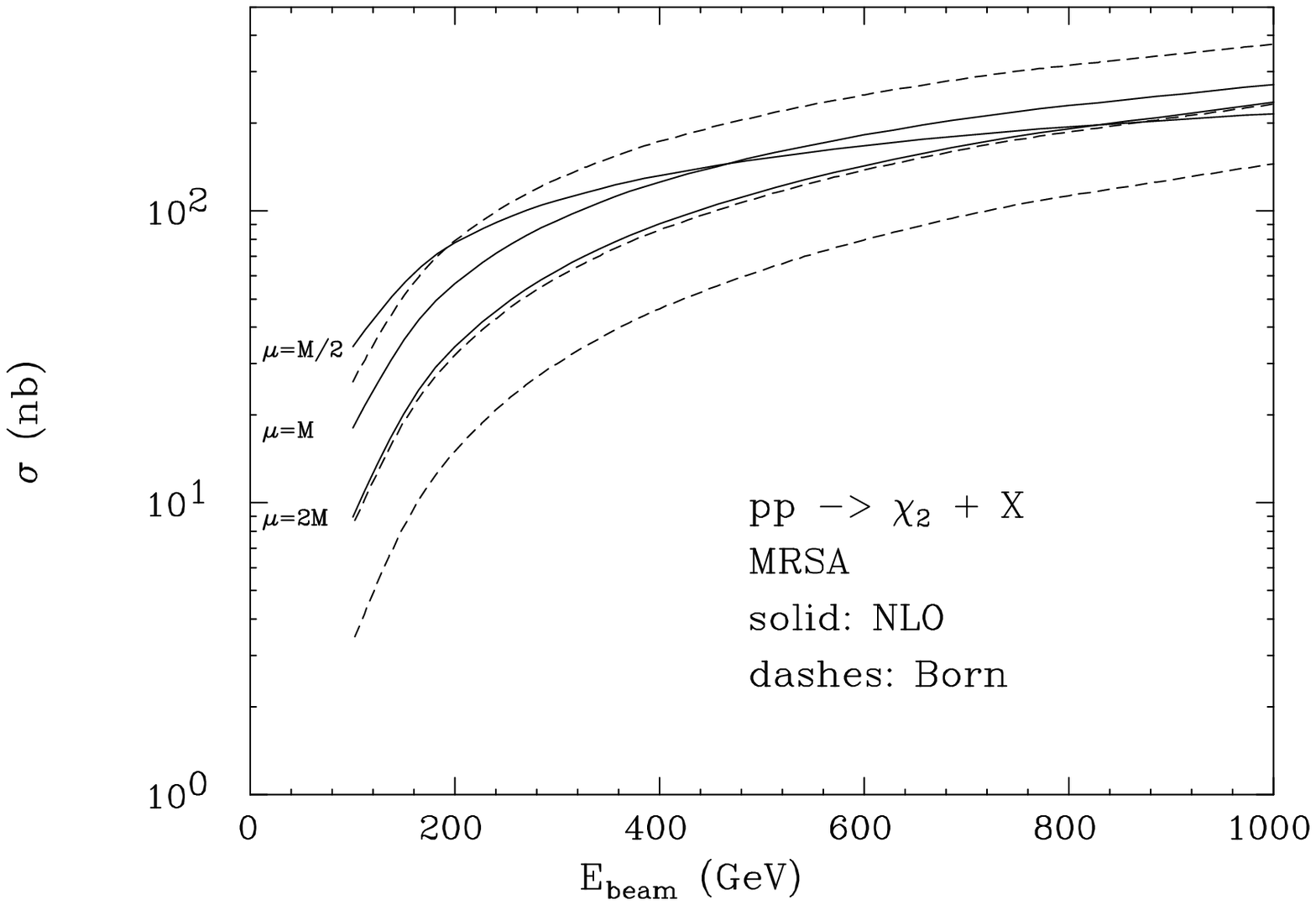,width=0.49\textwidth,clip=}
            \epsfig{figure=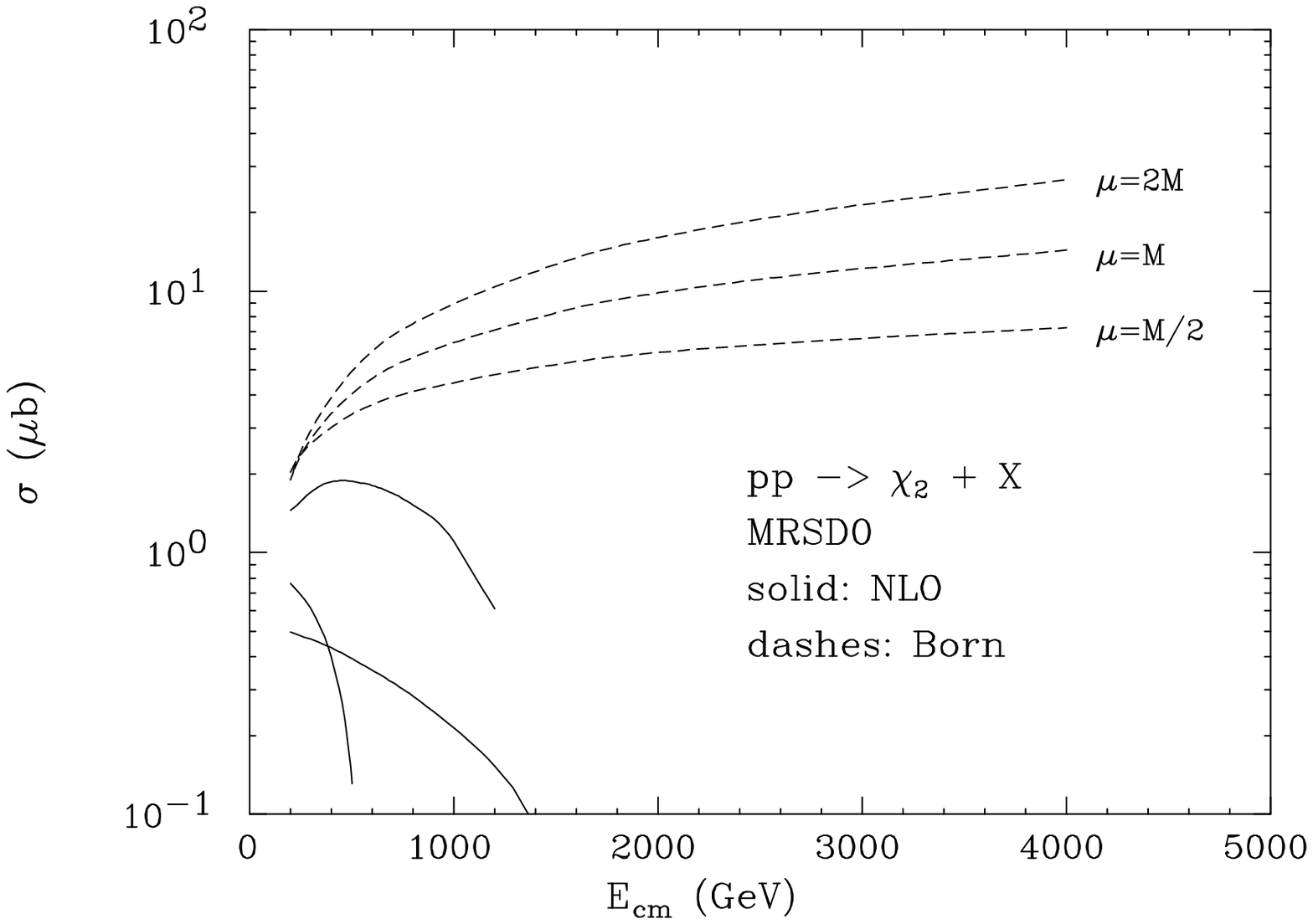,width=0.49\textwidth,clip=}}
\ccaption{}{ \label{fig:fxtchi}                 
Total cross section for $pp\to \chi_2X$ as a function of beam energy for
fixed-target collisions (a), and as a function of $\sqrt{S}$ for 
collider configurations (b).}
\end{figure}

The partonic cross sections described in the previous section can be used to
obtain total cross sections in hadronic collisions. For illustration, I will
consider here the case of $pp$ collisions.
In fig.~\ref{fig:fxtchi}(a) I present the results for
$\chi_{c,2}$ production at fixed-target energies, comparing the NLO to the LO
predictions. I use the MRSA~\cite{Martin94} PDF set, and show results for
three different scale choices. As can be seen, the NLO    
calculation significantly reduces the scale dependence of the LO result. The
size of the K-factor depends on the scale chosen, as well as on the beam
energy. The same distributions, plotted as a function of $\sqrt{S}$ in the
energy domain of the Tevatron collider, are given in fig.~\ref{fig:fxtchi}(b).
I chose here the MRSD0~\cite{Martin93} set of PDFs.
The results are extremely disappointing: not only have the NLO cross sections a
very strong scale dependence, but they also become negative for sufficiently
large $\sqrt{S}$. What is the origin of this behaviour, which makes the
perturbative evaluation of the cross sections totally unreliable?

There are at least two problems\footnote{Most of the remarks which follow have
already been made by G. Schuler in his '94 review~\cite{Schuler94}. Schuler at
the time had available the full NLO corrections to $\eta$ production, as well
as the leading small-$x$ behaviour of the $\chi$ cross sections. It is a pity
that those remarks have passed almost unnoticed in the community!}.
The first one is that the virtual corrections
are very large and negative. The large universal term $-C_A \as \pi/3 \sim 1$,
proportional to $\delta(1-x)$, is only in part                                 
cancelled by the state-dependent coefficient $A$ (see
eqs.~(\ref{eq:B0},\ref{eq:B2})). This indicates that two-loop corrections
coming from the square of the 1-loop matrix elements are likely to be 
large\footnote{Although the term
$-\as/\pi\,C_A/3\,\pi^2$ is universal and directly linked to the IR behaviour
of the real emission diagrams in a $gg\to X^{[1]}$ process, with $X^{[1]}$ an
arbitrary colour-singlet state, we have no argument suggesting that it should be
exponentiated. Other $\pi^2$ terms arise from the virtual corrections, not all
of them universal. Understanding which (if any) of them exponentiates requires
more work, which we have not done so far.}.
The second problem is that, after subtraction of the collinear singularities,
the \oacube\ corrections to both $gg$ and $qg$ processes tend to a negative
constant in the $x\to 0$ limit:                                    
\ba                    
 	\sigma_{NLO}^{J,(gg)}(x) &\stackrel{x\to 0}{\longrightarrow}&
	2C_A \asopi \sigma_{Born} \times \left( \log{\frac{m^2}{\muf^2}} -
                                                     C_J\right) \; , \\
 	\sigma_{NLO}^{J,(qg)}(x) &\stackrel{x\to 0}{\longrightarrow}&       
	\frac{C_F}{2C_A}\sigma_{NLO}^{J,(gg)}(x) \; ,
\ea                                                
where $C_J$=43/27 and 53/36 for $J=0$ and $J=2$, respectively.
There is nothing wrong in principle with these cross sections turning negative
in the small-$x$ region, as what is subtracted is partly returned to the gluon
density via the evolution equations. 
However in this particular case two things happen:
\begin{itemize}
\item the standard DGLAP evolution might not be adequate, as $x\ll 1$ at
collider energies.
\item the factorization scale (of the order of the charmonium mass) is very
close to the scale at which the input PDF is measured or parametrized, and
there is therefore no room for evolution (\ie  resummation of large logs). The
cross sections will therefore critically depend on the assumed shape of PDFs.
\end{itemize}
To illustrate the second point, let us assume for
simplicity that we can approximate the inelastic part of the
quarkonium cross section by its small-$x$ behaviour, in such a way that:
\be                                                                     
	\sigma_{NLO}(x) \; \sim \; A \delta(1-x) - \asopi C \; ,
\ee                               
(we left out irrelevant overall constants) and let us assume that we can
parametrize the gluon density with the following form:                  
\be
	G(x) \;=\; \frac{1}{x^{1+\delta}} \; ,
\ee
with $0<\delta<1$. 
It is then easy to show that the total cross section has the following
behaviours, depending on the value of the parameter $\delta$:
\be
\sigma_{NLO} \; = \; \sigma_{Born} \times \left\{ 
          \begin{array}{ll}
          A - C \asopi \log\frac{S}{m^2}  &  
                              \mbox{ if $\delta\log S/m^2\ll 1$} \\
          A - C \asopi \frac{1+\delta}{\delta} &               
                              \mbox{ if $\delta\log S/m^2\gg 1$} \\
          \end{array}                      \right.
\ee                                               
If the input gluon density $G(x)$ is not sufficiently steep (\ie if
$\delta\log S/m^2\ll 1$), very large logarithms $[\as\log S/m^2]^n$ will
appear at all orders of perturbation theory, and will need to be
resummed~\cite{smallx}, or accounted for by corrections to the DGLAP evolution.
If $\delta$ instead has a value of the order of 0.3--0.5, typical of the most
recent fits to HERA data, no large logarithmic corrections appear.
                        
\begin{figure}                                               
\centerline{\epsfig{figure=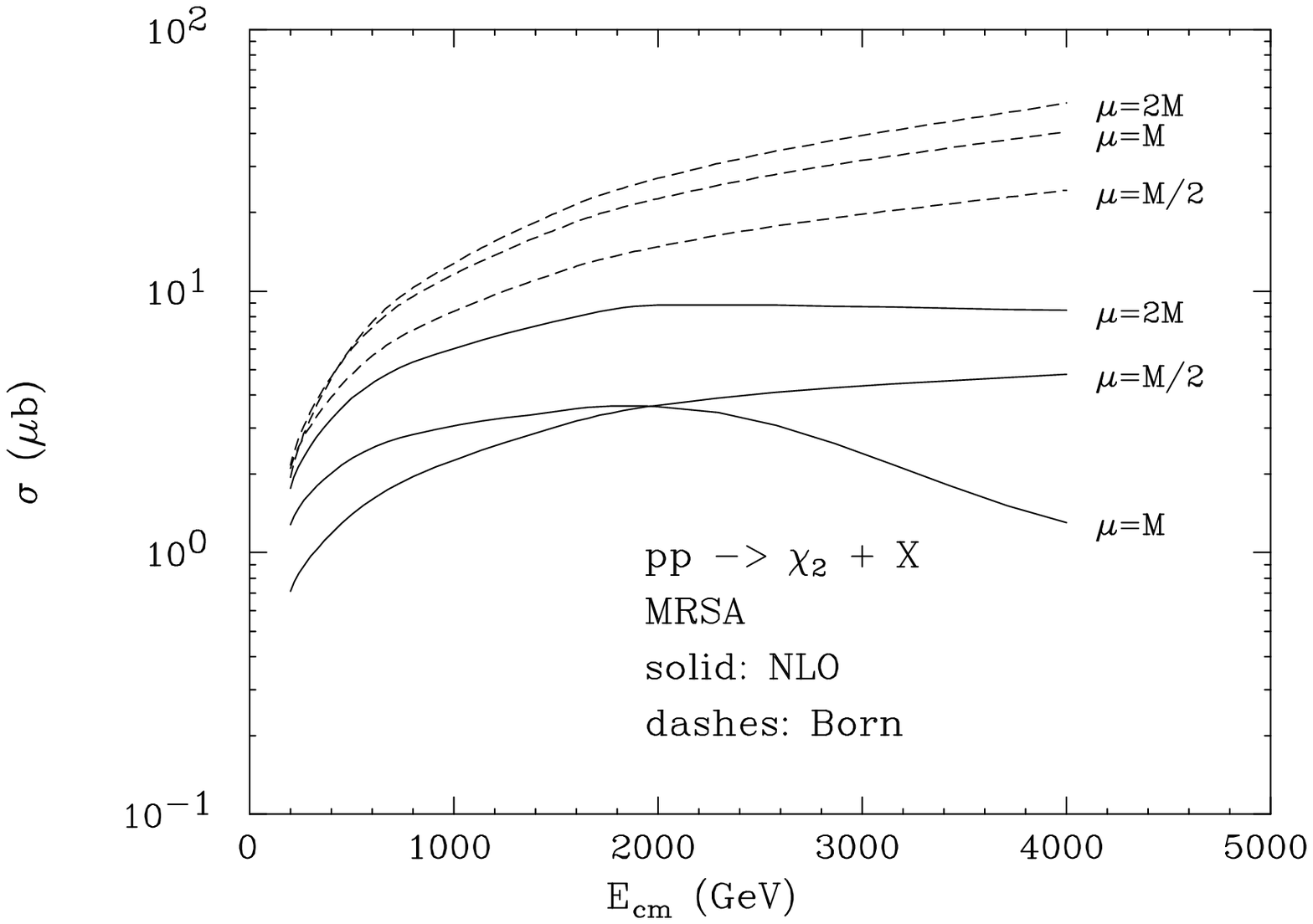,width=0.49\textwidth,clip=}
            \epsfig{figure=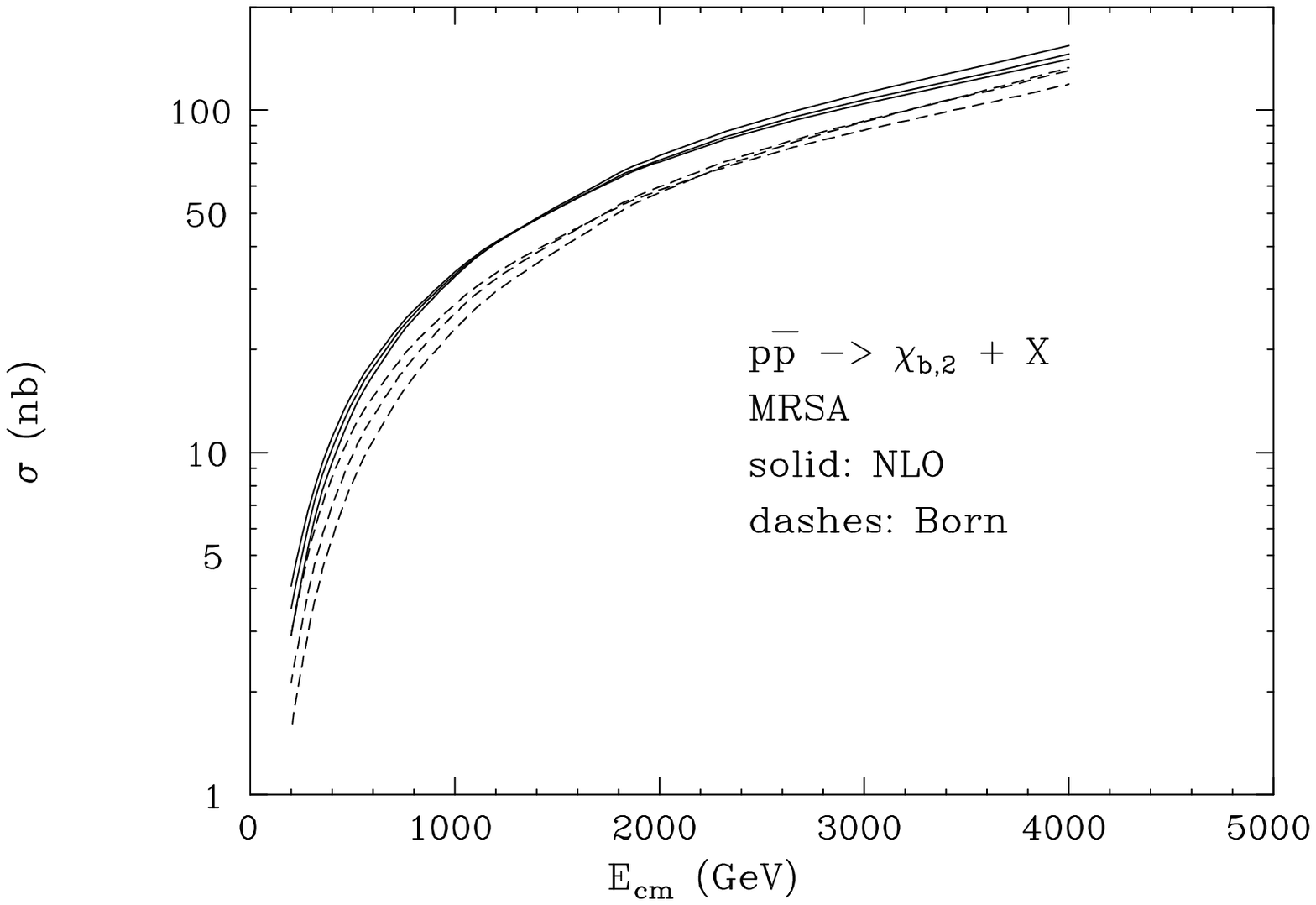,width=0.49\textwidth,clip=}}
\ccaption{}{ \label{fig:colchiA}                 
Left: total cross section for $pp\to \chi_{c,2}X$ as a function of 
$\sqrt{S}$ with PDF set MRSA. 
Total cross section for $pp\to \chi_{b,2}X$ as a function of 
$\sqrt{S}$ with PDF set MRSA. }      
\end{figure}                          
As an example of how a different choice of PDF can change things, I present in
fig.~\ref{fig:colchiA}(a) the $\chi_{c,2}$ cross sections obtained using the PDF
set MRSA, for which the input gluon density is steeper than $1/x$. The
NLO cross section remains now positive over a much larger range of $\sqrt{S}$.
Nevertheless the scale dependence is still so large that I would sadly conclude
that no predictive power is available at NLO for total $\chi_c$ production
cross sections at energies $\sqrt{S}$ larger than few hundred GeV. The
situation is significantly better in the case of bottomonium states, $\chi_b$,
shown in fig.~\ref{fig:colchiA}(b). In this case the inclusion of NLO
corrections significantly reduces the scale dependence of the LO result.
                                                                        
Needless to say, no conclusion on the behaviour of the charmonium cross
sections at large \pt\ can be reached from the previous study , since at large
\pt\ additional higher-order diagrams need to be calculated to achieve a NLO
accuracy, and since the range of $x$ and the scales probed are significantly
different than those explored in the total cross section calculation. No full
NLO calculation is currently available for quarkonium production in hadronic
collisions at non-zero \pt. Even the simplest case of $^1S_0$ production,
although irrelevant phenomenologically given that no data exist, might provide
an interesting theoretical insight if it were available. I would put this
calculation very high on the list of things to be done!
                                                                        
\section{Concluding Remarks}
In this final section I would like to address a few additional issues
related to the understanding of quarkonium production at higher orders in
perturbation theory:
\begin{itemize}
\item the exclusive structure of final states in quarkonium production via the
colour-octet mechanism and
\item the approximations involved in the use of the fragmentation functions for
production at large \pt.
\end{itemize}
As will appear from the following discussion,
the two issues are not entirely separated.

It is generally accepted by now that the proper treatment of the hierarchy of
higher Fock states is most likely the solution to a large fraction of the
puzzles present in quarkonium production data\footnote{That this is possibly
true for other, more exotic, aspects of quarkonium physics as well, has been
argued in the past and during this meeting by Brodsky~\cite{Brodsky96}}.   
It is also clear, however, that                                            
a complete understanding of the full dynamics of the interactions involving
colour-octet states (or, more generally, higher Fock states) is still missing.
To which extent this ignorance can affect our capability to perform quantitative
predictions for production rates is, in my view, unclear. To give an example,
let me consider charmonium production via the colour-octet mechanism at large
\pt. In this case, we believe that production is dominated by fragmentation
contributions, with a high-\pt\ gluon turning into some colour-octet state
($O$), that will then undergo a nonperturbative transition to 
a given, colour-singlet, onium state ($S$). This final step is
usually parametrized by assigning a probability, proportional to a well defined
nonperturbative matrix element, and by assuming that $S$
will carry all of the energy of the parent $O$. The               
nonperturbative transition $O\to S$ should however be thought of as an
inclusive process, $O\to S+X$, since one (or more) gluons need to be radiated.
In the standard approach, these gluons are assigned zero energy. In practice,
however, we know that these gluons cannot carry zero energy, because they will
have to materialize into some hadron, say into pions. 
In the rest frame of $O$, it is reasonable to assume that the energy of these
gluons will at least be of the order of $\Lambda_{QCD}$, \ie a number of the
order of, say,  300~MeV.  In we consider as an example the transition
$^3S_1^{[8]} \to ^3S_1^{[1]}$, believed to be responsible for the large
$J/\Psi$ rate measured at the Tevatron, there should be at least two gluons
emitted. In the laboratory frame, and for production at large \pt,
the ratio of the energy carried by these
gluons and the energy carried by the $^3S_1^{[1]}$ will be equal on average
to the ratio  of their masses, namely:
\be
	\frac{E_g}{E_\psi} \; = \; \frac{2\Lambda_{QCD}}{M_\psi} \sim 0.2\; .
\ee                                                                          
As a result, the actual energy fraction $z$ of the colour-octet state carried by
the final $J/\Psi$ will be around 0.8. So the fragmentation function instead of
peaking at $z=1$ will peak at $z=0.8$. Once convoluted with the \pt\ spectrum
of prompt gluons, assuming a behaviour like $d\sigma/d\pt \sim 1/\pt^n$,
this change will induce a change in the production rate
at a given \pt\ of the order of $-n\times 0.2$. For a typical value of $n=4$,
this is a $-80\%$ correction.        
It is important to point out that this is not a higher-twist phenomenon, 
in the sense that the effect is not reduced by going to larger \pt.
In order to make a more accurate prediction of the high-\pt\ production rate,
it is therefore essential to improve the understanding of the colour-bleaching
mechanism\footnote{As an aside, we remark here that this understanding could
also lead to some specific and testable predictions. For example, the interplay
between quantum numbers and phase space might lead to interesting selection
rules on the possible sets of light hadrons produced in the $O\to S$
transition.}
Notice also that this problem is not totally unrelated to the
problem of connecting nonperturbative matrix elements ``measured'' in
quarkonium decay via colour-octet states to the nonperturbative matrix elements
needed in the evaluation of production cross sections.

Let me now touch on one more issue related to fragmentation. The standard
approach consists in determining the perturbative boundary condition for the
evolution of the fragmentation function by using the following
relation~\cite{Braaten93}:
\be
	D_0(z,m^2) \= \int_{m^2}^{\infty} \frac{ds}{s} \; d(z,s) \;,
\ee          
where $d(z,s)$ is the probability that a gluon of virtuality $s$ decays to a
$J/\Psi$ carrying longitudinal momentum fraction $z$ in the infinite-momentum
frame. The evolution of the fragmentation function $D(z,q^2)$ is then given by
the standard DGLAP evolution equation:
\be
	\frac{\partial}{\partial\log\mu^2} D(z,\mu^2) \=
       \asopi \int_z^1 \; \frac{dy}{y} P_{gg}(z/y)  D(z,\mu^2) \;.
\ee                                                              
The problem with this equation~\cite{frag}
is that it does not respect the phase-space
constraint $D(z,\mu^2)=0$ for $z < m^2/\mu^2$.  The implementation of 
this  constraint would slow down the evolution of the fragmentation function 
by delaying the depletion of the large-$z$ fragmentation region.  Since the 
spectrum of gluons falls rapidly with $p_T/z$, a proper 
treatment of the large-$z$ region can have a significant effect on the cross 
section.  
A more accurate treatment would be to use the following 
equation:
\be \label{nhap}
\mu^2 \frac{\partial}{\partial \mu^2} D(z,\mu^2) \;=\;
d(z,\mu^2) \;+\; \frac{\alpha_s}{2\pi}
\int^{1}_{z} \frac{dy}{y} \; P_{gg}(y)
\; D(z/y, y \mu^2)  ,
\ee
%[[ I THINK THE 2ND ARGUMENT OF $D$ SHOULD BE $z \mu^2/y$ ]] 
together with the boundary condition $D(z,\mu^2=m^2) = 0$.  This 
evolution equation respects the phase space constraint, as can be easily 
checked. 
A consistent calculation done using NLO fragmentation functions should use this
evolution, instead of the naive one.

As a final point, I would like to present a simple proposal for how to
describe the exclusive structure of the final state in 
quarkonium production via fragmentation through a colour-octet state.
After generation of a hard final-state gluon, let the gluon
shower evolve, until it generates a $c\bar c$ pair. 
Allow then the $c\bar c$ pair to evolve.
If additional gluons are emitted, we can assume that 
no quarkonium state should be produced.
This would be in fact a $1/N_c$ suppressed process. Given that the
emitted gluons are perturbative, the only way to correctly calculate the
probability that the colours will recombine into a singlet state after gluon
emission is by using the colour-singlet matrix elements. If no additional
gluons are emitted, then consider the invariant mass of the pair. If it is
below a given value (to be parametrized by the BBL factorization scale, the
scale at which NRQCD matrix elements are separated from the perturbative
ones) then we can assume that the $c\bar c$ pair will be converted into a 
colour-singlet $J/\Psi$ plus two gluons, with a probability proportional to the
NRQCD matrix element (this matrix element depends on the factorization scale,
so at the end the factorization scale dependence should cancel between the
choice of the phase space boundary and the transition probability).  This
transition is equivalent to what is done in the cluster model for
hadronization~\cite{Marchesini88}, where at the end of the       
perturbative evolution each gluon is split into a $q\bar q$ pair. The energy and
angular distribution of the two gluons can just be taken to be given by the
3-body phase space for transition of the $c\bar c$ pair into the $J/\Psi g g$
final state. Colour lines can be drawn between these two gluons and the
rest of the shower, so that hadronization can take place (say via the
cluster model itself). From the point of view of the algorithm
efficiency, one could use the forced evolution a' la Mike
Seymour~\cite{Seymour94} to
always get a $c\bar c$ pair at the end of each shower. All of the
MC inefficiency is
then related to the invariant mass distribution of the pair, which will often
be above the factorization scale threshold.                                   
                
The first attempt to produce $J/\Psi$ states via the colour-octet mechanism in
a full shower MC, using however a different approach than what suggested
above, has recently been presented by Ernstr\"om and
L\"onnblad~\cite{Ernstrom96}. This calculation allows to make definite
predictions for the structure of the gluon-jet which accompanies the $J/\Psi$.
Comparisons of these predictions with data will certainly help improving  
our understanding of this important aspect of the production dynamics.
                                                                      
It is worth keeping in mind that all the effects discussed in this section can
lead to significant changes in the shape of the \pt\ distribution of $J/\Psi$s 
observed at the Tevatron. In addition to the factors described here, one should
consider of course the effect of higher-order corrections due to multiple-gluon
emission from the initial-state, and the systematic uncertainties due to the choice of the input gluon
densities and due to the chosen value of \as. All of these effects
will also induce a smearing of the \pt\ spectrum, and could therefore influence
dramatically the extraction of the nonperturbative parameters for the different
channels which contribute to $J/\Psi$ production at the Tevatron.

\end{document}